\documentclass[authoryear]{FLO_v1}%
\usepackage{graphicx}
\usepackage{upgreek}
\usepackage{multicol,multirow}
\usepackage{amsmath, amssymb, amsfonts}
\usepackage{mathrsfs}
\usepackage{amsthm}
\usepackage{epsfig}
\usepackage[figuresright]{rotating}
\usepackage{appendix}
\usepackage[authoryear]{natbib}
\usepackage{ifpdf}
\usepackage[T1]{fontenc}
\usepackage{newtxtext}
\usepackage{newtxmath}
\usepackage{textcomp}
\usepackage{xcolor}
\usepackage{epsfig, psfrag, epstopdf}
\usepackage{float}
\usepackage{tikz}

\usepackage{physics, tikz}
\usepackage[colorlinks,allcolors=blue]{hyperref}
\definecolor{jourcolor}{cmyk}{1,0.57,0.01,0.38}
\hypersetup{
    colorlinks,%
    citecolor=jourcolor,%
    filecolor=jourcolor,%
    linkcolor=jourcolor,%
    urlcolor=jourcolor
}

\newcommand*\circled[1]{\tikz[baseline=(char.base)]{
\node[shape=circle,draw,inner sep=1.2pt] (char) {#1};}}

\theoremstyle{definition}

\articletype{RESEARCH ARTICLE}

\citearticle{Tomar et al.}

\begin{document}

\title[Wall Shear Stress Generated by a Bernoulli Pad: Experiments and Numerical Simulations]{Wall Shear Stress Generated by a Bernoulli Pad: Experiments and Numerical Simulations}

\author[Tomar, et al.]{Anshul S. Tomar$^{1}$, Shaede Perzanowski$^{1}$, Ricardo Mejia-Alvarez$^{1}$, Ranjan Mukherjee$^{1,*}$, Aren Hellum$^{2}$ and Kristina Kamensky$^{2}$}
\address[1]{Michigan State University, East Lansing, MI 48824, USA}
\address[2]{Naval Undersea Warfare Center, Newport, RI 02841, USA}
\corres{*}{Corresponding author. E-mail:
\emaillink{mukherji@egr.msu.edu}}


%\date{\textbf{Received:} XX 2020; \textbf{Revised:} XX XX 2020; \textbf{Accepted:} XX XX 2020}

\abstract{Bernoulli pads generate locally large wall shear stresses on workpieces, which can be used for cleaning, but may also damage delicate surfaces. This work presents direct measurements of the wall shear stress using constant-temperature anemometry for the first time. A hot-film sensor was calibrated in the laminar and turbulent flow regimes using a purpose-built water flow channel. The calibrated sensor was then flush-mounted onto a smooth surface and a Bernoulli pad was traversed over the sensor and wall shear stress data were acquired. Numerical simulations of the flow field were also performed; they accurately predicted the maximum shear stress near the jet corner but over-predicted at large radii.}

\maketitle

\begin{boxtext}

\textbf{\mathversion{bold}Impact Statement}

Biological organisms accumulate on the hulls of ships, leading to increased energy consumption, higher maintenance costs, and the transport of invasive organisms. Fouling-release coatings which allow organisms to be removed by shear are prone to mechanical damage and are not effective when the ship is stationary. A Bernoulli pad confines a high-speed jet of fluid to a small gap, producing a normal force which maintains the gap and wall shear which can be used to clean the hull. This flow field is separated near the jet core, transitioning, and highly confined, which makes it difficult to study experimentally and challenging to model. In particular, the shear force produced by a pad as a function of radial position has not been measured. In this work, we measure the wall shear stress of a Bernoulli pad using hot-film anemometry. We also compare these results to numerical simulations, and discuss shortcomings observed in the modeling results.

%Biofouling mitigation from ship hulls is an essential process of removal of biological organisms that accumulate on the hull surface. The colonization of biological organisms on the hull surface can lead to increased energy consumption, higher maintenance costs, and damage to the alien ecosystem. Biofouling mitigation from ship hulls can be obtained using a non-contact Bernoulli pad, which utilizes the shear stress produced to clean the surfaces. The unique flow physics of Bernoulli pads, involving flow separation, transition, and flow through a very small gap, makes it very difficult to study them experimentally. The available literature in this field consists of numerical simulations and experiments to quantify the normal forces produced by the Bernoulli pad but lacks shear stress quantification using experiments. This study aims to fill this important literature gap by measuring the maximum wall shear stress, which can be used to validate the numerical models and help utilize numerical simulations for more generic design optimizations of the Bernoulli pad for biofouling mitigation.

\end{boxtext}

\section{Introduction}

A Bernoulli pad is conventionally used to pick and place objects without contacting them (\cite{bib1, bib2}). The pad is proximally located to an object or a workpiece, and axial flow through the center of the pad impinges on the workpiece and is deflected radially outward. The center of this impingement region is a stagnation point, where pressure is the highest. As the flow is deflected radially outward in the impingement region, the gage pressure decreases gradually but remains positive as it approaches the entrance to the pad's gap.  As such, this impingement region tends to repel the workpiece. The drastic increase in velocity due to the cross-section reduction at the entrance to the pad's gap induces a pressure drop down to vacuum levels. As the flow expands radially outward inside the pad's gap, the intensity of this vacuum reduces gradually. The vacuum present in the pad's gap induces an attractive force on the workpiece. The balance between the repulsive force from the impingement region and the attractive force from the pad's gap tends to fix the gap at a constant value. Any effort to take the pad from this equilibrium configuration is met with a resistive force. The effect of the gap on the nature of the normal force (attractive or repulsive) has been widely discussed in the literature (\cite{bib3}). The change in the nature of the force gives rise to both stable and unstable equilibrium configurations (\cite{bib4}) but the presence of the stable equilibrium configuration allows the Bernoulli pad to be used for pick and place operations in industry (\cite{bib5, bib6, bib7}). \

In addition to normal forces, shear forces are generated by the flow field between the pad and the workpiece (\cite{bib8}). Flow-induced shear forces generated by a Bernoulli pad have found the application of non-contact biofouling mitigation of ship hulls (\cite{bib9}). A Bernoulli-pad device can be used to remove biofouling while the ship is at port, in both wet and dry docking conditions, and can leverage the benefits of fouling release coatings (\cite{bib10}). The organisms which make up the biofouling colonies can also be altered by repeated application of fluid shear stress (\cite{bib40}). The abrupt change in the direction of flow in a Bernoulli pad introduces separation and recirculation near the neck of the pad (\cite{bib11}). The separation and constriction of the flow results in large magnitude shear stresses on the workpiece, which are essential for the application of bio-fouling mitigation.\

In our previous work (\cite{bib4}), we have used computational fluid dynamics simulations to develop a better understanding of the flow physics associated with a Bernoulli pad, including the location and magnitude of the maximum shear stress. It was found that the magnitude of wall shear stress is maximum at the belly of the recirculation region. Resolving the region of flow separation is computationally expensive, which makes it challenging to predict the wall shear stress accurately. The numerical results need validation, and for the first time, in this paper a constant temperature hot-film anemometer is used with water as the working fluid to measure the wall shear stress. Measurement of the shear stress profile will aid in the design of Bernoulli pads as hull grooming devices and help create cleaning schedules as cleaning efficacy is dependent on the organism and time between grooming (\cite{bib12}).\

Over the last few decades, various methods for wall shear stress measurements have been developed and reviewed in the literature (\cite{bib13, bib14, bib15, bib16}). A method that combines good spatial resolution with the ability to work in water and be implemented in a confined geometry with poor optical access is hot-film anemometry. Hot-film sensors have been used to detect flow separation and detection of transition from laminar to turbulent flow - see (\cite{bib17, bib18, bib19}), for example. In the present work, we use a flush-mounted hot-film sensor for measurement of wall shear stress generated by a Bernoulli pad. The main principle behind this technique is to correlate heat transfer from the sensor with wall shear stress. To this end, the sensor needs to be calibrated under known wall shear conditions. Under a Bernoulli pad, the maximum wall shear occurs in a very narrow region close to the neck of the pad, but decays rapidly at larger radii (\cite{bib4, bib20}). This large variation makes it challenging to investigate the wall shear stress experimentally.\

Experimental investigations with Bernoulli pads have been reported in the literature. \cite{bib3}, \cite{bib21}, \cite{bib22}, and \cite{bib23}, for example, conducted experiments to understand the various factors that affect the wall-normal forces. A majority of the investigations in the literature have focused on wall-normal forces and used air as the working fluid. To the best of our knowledge, the only work with water as the working fluid was reported by \cite{bib21} and \cite{bib24}. The work by \cite{bib21} focused on the normal force and pressure distribution generated by the flow field, whereas Particle Tracking Velocimetry (PTV) experiments were conducted to measure the velocity components of the flow field in \cite{bib24}. PTV experiments do not provide reliable measurements close to the wall and cannot be used to accurately measure wall shear stress. Hence, the wall shear measurements presented in this work fill an important gap in the literature. It should be mentioned that experimental studies on radial suction flow and its effect on water vortex unit was carried out in \cite{bib25}. Although water is used as the working fluid, the work focused on normal suction forces and not shear force measurements.\

This paper is organized as follows. A brief review of channel flow and the working principle of hot-film sensors are provided in Section \ref{sec2}. The hot-film sensor is calibrated using fully-developed channel flow; an analytical solution for wall shear in channel flow is presented in Section \ref{sec2} and the procedure for calibration of the hot-film sensor is presented in Section \ref{sec3}. The experimental setup for wall shear measurements is described in Section \ref{sec4}. Section \ref{sec5} provides experimental results and compares them with results obtained from numerical simulations. Concluding remarks are provided in Section \ref{sec6}.\

\section{Background}\label{sec2}

\subsection{Analytical solution for wall shear in channel flow}\label{sec22}

\begin{figure*}[t!]
 \begin{center}
% \psfrag{A}[][]{\small{$x$}}
% \psfrag{B}[][]{\small{$z$}}
% \psfrag{C}[][]{\small{$y$}}
% \psfrag{D}[][]{\small{$L$}}
% \psfrag{E}[][]{\small{$w$}}
% \psfrag{F}[][]{\small{$h$}}
% \psfrag{M}[][]{\small{$c$}}
% \psfrag{N}[][]{\footnotesize{$\circled{2}\!$ {\scriptsize and} $\!\circled{3}$}}
% \psfrag{P}[][]{\footnotesize{$\circled{1}$}}
% \psfrag{V}[][]{\footnotesize{$\circled{8}$}}
% \psfrag{R}[][]{\footnotesize{$\circled{4}$}}
% \psfrag{S}[][]{\footnotesize{$\circled{5}$}}
% \psfrag{T}[][]{\footnotesize{$\circled{6}$}}
% \psfrag{U}[][]{\footnotesize{$\circled{7}$}}
% \includegraphics[width=5.1in]{Fig1.eps}
  \includegraphics[width=5.1in]{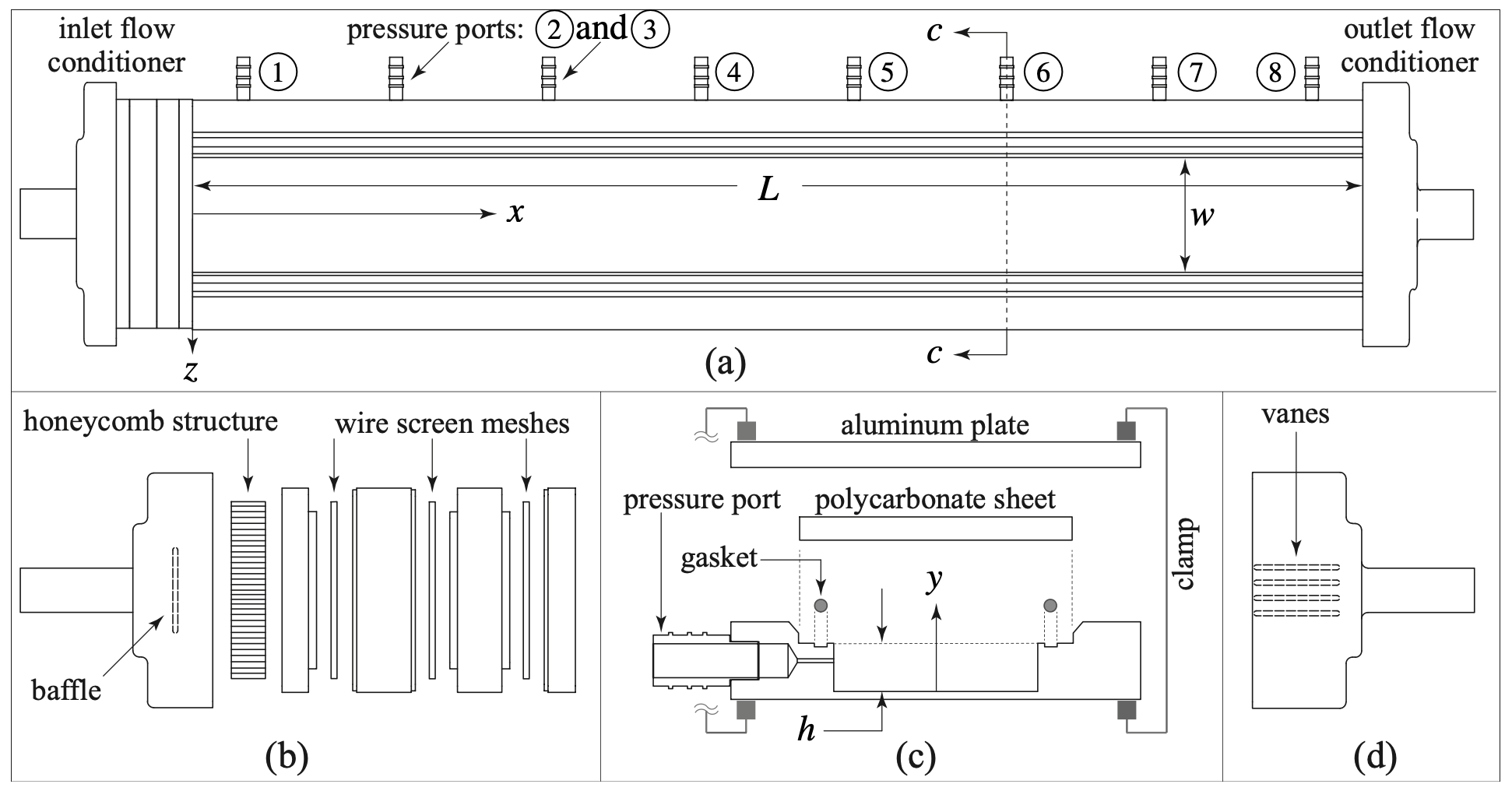}
 \caption{\small{Channel setup used for calibration of the hot-film sensor; figures are not drawn to scale: (a) top-view of the channel showing the eight pressure ports, (b) exploded view of the inlet flow conditioner, (c) exploded view of section c-c of channel with gasket, polycarbonate sheet, aluminum plate, and clamps (d) outlet flow conditioner showing internal vanes.}}
 \label{Fig1}
 \end{center}
 \end{figure*}

For two-dimensional steady-state fully developed channel flow - see Fig.\ref{Fig1}, the Navier-Stokes equation for the streamwise velocity becomes:
\begin{equation*}
\mu \frac{d^2 u}{d y^2} = \frac{dp}{dx}
\end{equation*}

\noindent where $p = p(x)$ is the pressure, $u=u(y)$ is the streamwise velocity, and $\mu$ denotes the dynamic viscosity. On integrating the above equation with respect to $y$ once,  and substituting the symmetry condition $d u/d y = 0$ at $y = h/2$, we get:
\begin{equation}
\label{eq1}
\mu\frac{d u}{d y} = \left(y - \frac{h}{2}\right)\frac{dp}{dx}
\end{equation}

\noindent Note that $h$ is the channel's height. This result does not assume any particular flow regime, laminar or turbulent. From the above result, the shear stress $\tau$ can be written as:
\begin{equation*}
\tau \triangleq \mu \frac{d u}{d y} = \left(y - \frac{h}{2}\right) \frac{dp}{dx}
\end{equation*}

\noindent At the wall, where $y = 0$, the wall shear stress $\tau_{\rm w}$ is then given by:
\begin{equation}
\label{eq2}
\tau_{\rm w} = -\frac{h}{2}\frac{dp}{dx}
\end{equation}

\noindent which implies that the wall shear stress in channel flow can be determined by measuring the pressure gradient. The channel shown in Fig.\ref{Fig1} was constructed to provide a linear pressure gradient over a large distance, so that the flow is fully developed and $({dp}/{dx})$ can be confidently estimated.\

\subsection{Working principle of hot-film anemometry}\label{sec21}

\begin{figure}[b!]
\begin{center}
%\psfrag{A}[][]{\small{$0.75$ mm}}
%\psfrag{B}[][]{\small{$0.2$ mm}}
%\includegraphics[width=0.3\hsize]{Fig2.eps}
\includegraphics[width=0.3\hsize]{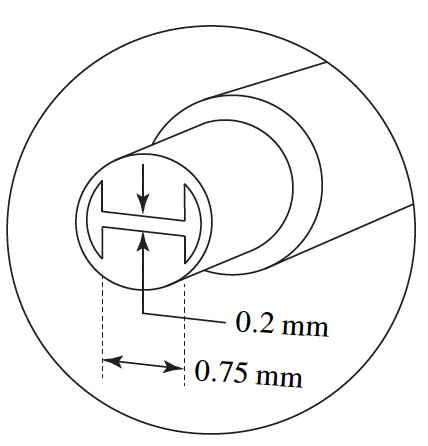}
\end{center}
\caption{\small{A schematic of the hot-film sensor 55R46 by Dantec Dynamics (\cite{bib30})}}
\label{Fig2} 
\end{figure}

Hot-film anemometry is used to measure the velocity and turbulence properties of fluid flows by measuring the heat dissipated due to convection. The hot-film sensor used in our experiments is shown in Fig.\ref{Fig2}. The H-shaped film at the tip of the sensor is a very thin electrical resistor, through which current is passed, generating heat. The heat transfer rate from the film into the fluid varies with flow velocity according to King's law (\cite{bib26}):

\begin{equation} \label{eq3}
q_{\rm conv} = a+b u^n, 
\end{equation}

\noindent where $a$, $b$, $N$ are constants, and $u$ is the characteristic velocity near the sensor. The heat generated by electric current is:
\begin{equation} \label{eq4}
q_{\rm gen} = \frac{E^2_{\rm a}}{R_{\rm f}}
\end{equation}

\noindent where $E_{\rm a}$ is the voltage applied across the film, and $R_{\rm f}$ is the film's electrical resistance. The resistance $R_{\rm f}$ is held constant using a Constant Temperature Aneomometer (CTA), and $E_{\rm a}$ is monitored using a Wheatstone bridge. At constant temperature, energy conservation yields $q_{\rm conv} =q_{\rm gen}$. Equations \eqref{eq3} and \eqref{eq4} can be combined and manipulated to yield the well-known relation:

\begin{equation} \label{eq8}
E^2_{\rm a} = A+B\tau_{\rm w}^n, 
\end{equation}

\noindent where $\tau_{\rm w}$ is the wall shear stress (\cite{bib27}). The calibration coefficients $A$, $B$, $n$ are determined by correlating the voltage output of the sensor with known wall shear stress in fully-developed channel flow. Because the heat transfer in laminar flow is significantly different than that found in turbulent flow, two sets of coefficients are required for flow fields which contain both regimes.\

\section{Calibration of Hot-Film Sensor}\label{sec3}

\subsection{Design of channel}\label{sec31}
\begin{figure*}[t!]
\begin{center}
%\psfrag{M}[][]{\footnotesize{$\circled{2}\!$ and $\!\circled{3}$}}
%\includegraphics[width=5.5in]{Fig3.eps}
\includegraphics[width=5.5in]{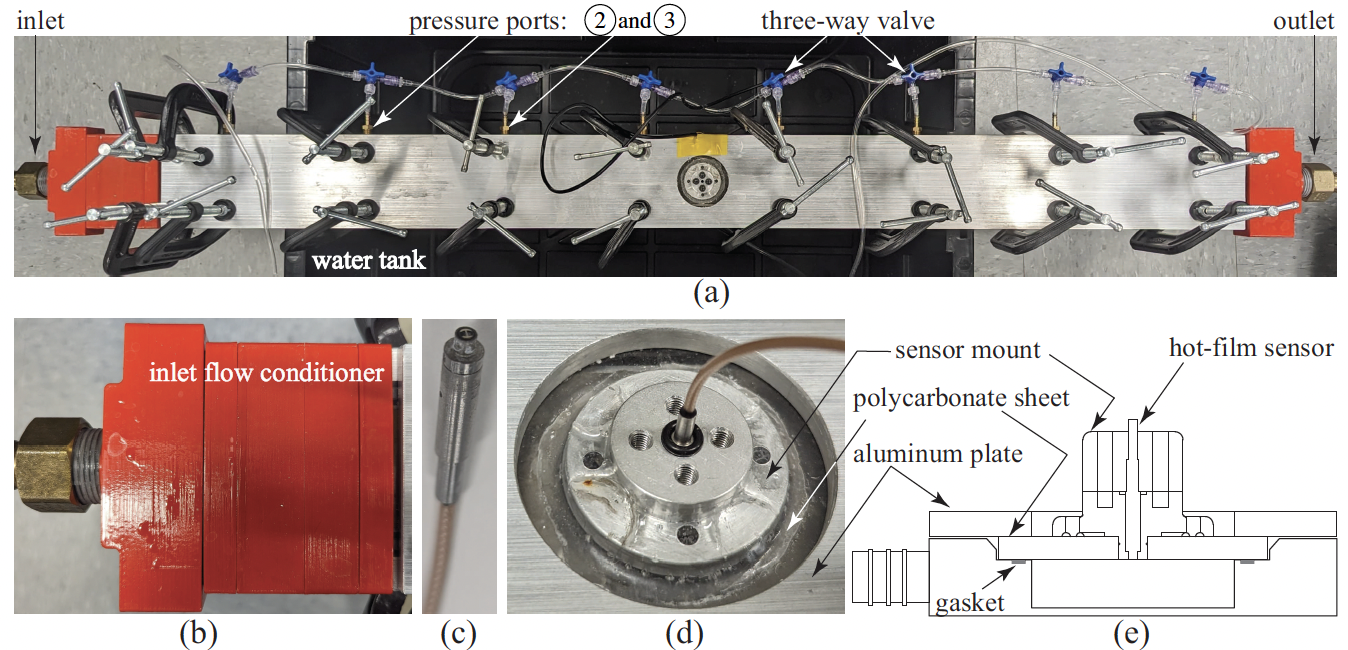}
\caption{\small{Assembled view of the channel setup in Fig.\ref{Fig1}: (a) top-view of the channel showing sensor mount (without sensor) and eight pressure ports (see Fig.\ref{Fig1}) connected via three-way valves, (b) inlet flow conditioner, (c) hot-film sensor, (d) magnified view of sensor mount with sensor, (e) sectional view of channel setup through the sensor mount and sensor.}}
\label{Fig3}
\end{center}
\end{figure*}

The channel used for calibration of the hot-film sensor is motivated by prior work reported in the literature (\cite{bib28}). A schematic of the channel is shown in Fig.\ref{Fig1}(a) and a sectional view through a pressure port is shown in Fig.\ref{Fig1}(c). The channel was fabricated using aluminum and its cross-section ($w = 50$ mm, $h = 5$ mm) was chosen to target a Reynolds number of ${\rm Re} = 10,\!000$  with the available pump.\

The length of the channel was chosen to be $L = 1,\!200$ mm to ensure that the flow would be fully-developed over a significant distance. The top of the channel is covered with a clear polycarbonate sheet and leakage is prevented through the use of a gasket along the length of the channel. An aluminum plate is placed over the polycarbonate sheet and clamps are used to apply uniform pressure on the gasket along the length of the channel to make it leak proof - see Fig.\ref{Fig1}(c) and Fig.\ref{Fig3}(a).\

A submersible utility pump was used to circulate water through the channel. The flow rate is controlled with a gate valve installed upstream of the channel's inlet. The flow is conditioned at the inlet of the channel with a series of baffles, followed by a honeycomb, and then a series of three meshes of decreasing hole size - see Fig.\ref{Fig1}(b) and  Fig.\ref{Fig3}(b). To reduce the developing length, the boundary layer is tripped with a coarse-grit sand paper strip at the inlet of the channel.  The outlet of the channel is connected to a flow conditioner with four internal vanes to minimize end-effects - see Fig.\ref{Fig1}(d).\
\begin{figure}[t!]
\begin{center}
%\psfrag{A}[][]{\small{$p$ (kPa)}}
%\psfrag{1}[][]{\footnotesize{$\circled{1}$}}
%\psfrag{2}[][]{\footnotesize{$\circled{2}$}}
%\psfrag{3}[][]{\footnotesize{$\circled{3}$}}
%\psfrag{4}[][]{\footnotesize{$\circled{4}$}}
%\psfrag{5}[][]{\footnotesize{$\circled{5}$}}
%\psfrag{6}[][]{\footnotesize{$\circled{6}$}}
%\psfrag{7}[][]{\footnotesize{$\circled{7}$}}
%\psfrag{8}[][]{\footnotesize{$\circled{8}$}}
%\includegraphics[width=0.5\hsize]{Fig4.eps}
\includegraphics[width=0.5\hsize]{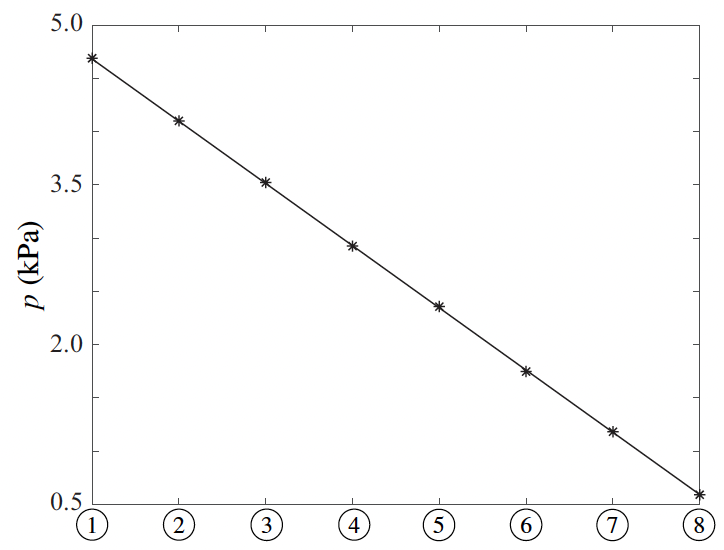}
\end{center}
\caption{\small{Pressure at the eight different ports of the channel (see Fig.\ref{Fig1}), computed based on pressure differential measurements relative to port $\protect\circled{1}$ and assignment of an arbitrary pressure to port $\protect\circled{8}$. Note that the straight line fit was obtained by using the data from ports $\protect\circled{4}$ though $\protect\circled{8}$.}}
\label{Fig4} 
\end{figure}

To measure pressure along the channel, eight pressure taps are placed on the side of the channel along its length. The ports are $150$ mm apart and drilled with a size of $0.8$ mm. The ports are connected by a three-way valve - see Fig.\ref{Fig3}(a), to a DP15 differential pressure transducer (a product of Validyne Engineering (\cite{bib29}), not shown. The pressure differential between ports $\!\circled{2}$ through $\!\circled{8}$ relative to $\!\circled{1}$ are recorded at steady state; the data is then used to compute the pressure at all the ports by assigning an arbitrary pressure to port $\!\circled{8}$ - see Fig.\ref{Fig4}. We observe a linear pressure gradient over the entire measurement section, which indicates that the flow is fully developed upstream of port $\!\circled{1}$. We mounted the hot-film sensor between ports $\circled{4}$ and
$\circled{5}$ on the polycarbonate sheet, as shown in Fig.\ref{Fig3}(a). A closeup view of the sensor mount is provided in Fig.\ref{Fig3}(d). A sectional view of the assembled channel setup through the sensor mount and sensor is shown in Fig.\ref{Fig3}(e).\

\subsection{Calibration procedure}\label{sec32}

\begin{figure}[t!]
\begin{center}
%\psfrag{A}[][]{\small{$E_{\rm a}^2$ (V$^2$)}}
%\psfrag{B}[][]{\small{$\tau_{\rm w}$ (Pa)}}
%\includegraphics[width=0.5\hsize]{Fig5.eps}
\includegraphics[width=0.5\hsize]{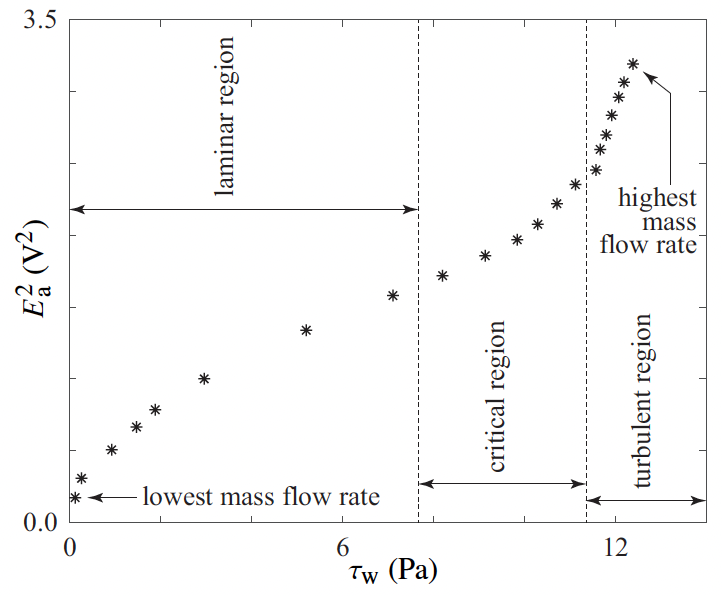}
\end{center}
\caption{\small{Calibration data showing the variation of $E_{\rm a}^2$ with $\tau_{\rm w}$}}
\label{Fig5} 
\end{figure}

The hot-film sensor 55R46, shown in Fig.\ref{Fig2} and Fig.\ref{Fig3}(c), was calibrated using the constant temperature anemometer MiniCTA 54T42, also a product of Dantec Dynamics (\cite{bib30}). The calibration was performed using 21 different flow rates through the channel; a gate valve was installed between the prime mover and the inlet of the channel to control the flow rate. For each flow rate, the pressure transducer and hot-film sensor measurements (voltages) were recorded. The pressure measurement (voltage) provides the pressure drop between ports $\circled{4}$ and $\circled{5}$; transducer calibration data is used to express it in Pa and the wall shear stress $\tau_{\rm w}$ is then computed using Eq.\eqref{eq2}. The variation of the square of the hot-film sensor voltage ($E_{\rm a}^2$) with the wall shear stress $\tau_{\rm w}$ is plotted in Fig.\ref{Fig5} with the objective of computing the calibration coefficients in Eq.\eqref{eq8}. It can be seen that the wall shear stress increases monotonically with increase in the mass flow rate. Also, the variation of $E_{\rm a}^2$ with $\tau_{\rm w}$ depicts three distinct behaviors corresponding to three distinct flow regimes: turbulent, critical and laminar. In the laminar and turbulent regimes, the data points show a concave downward trend with increase in wall shear stress; the trend is reversed and is concave upward in the critical regime.\
\begin{figure}[t!]
\begin{center}
%\psfrag{A}[][]{\small{$E_{\rm a}^2$}}
%\psfrag{B}[][]{\small{$\tau_{\rm w}^{0.4}$}}
%\psfrag{C}[][]{\small{$\tau_{\rm w}^{0.2}$}}
%\includegraphics[width=0.5\hsize]{Fig6.eps}
\includegraphics[width=0.5\hsize]{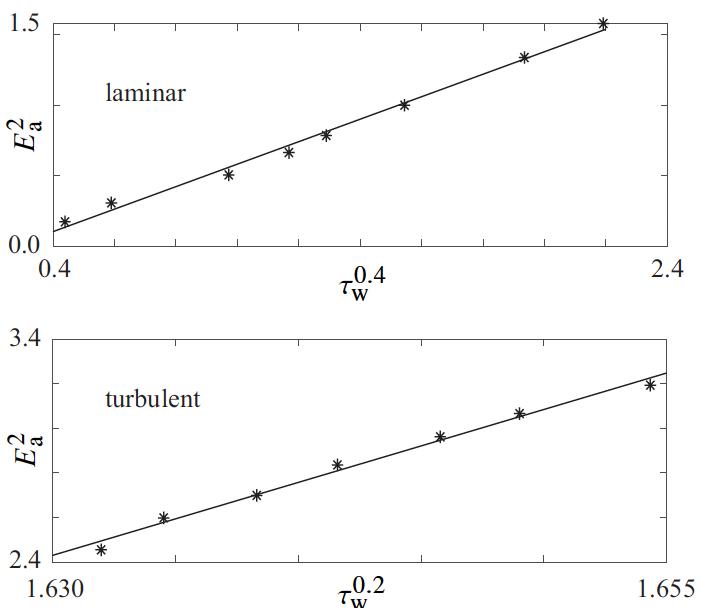}
\end{center}
\caption{\small{Linear calibration curves for laminar and turbulent regimes}}
\label{Fig6} 
\end{figure}

The data points corresponding to the laminar and turbulent regimes are used to find a linear fit between $E_{\rm a}^2$ and $\tau_{\rm w}^n$, where the value of the exponent $n$ should lie in the range $[0.1, 0.5]$ (\cite{bib27}). A linear fit is found by choosing $n=0.4$ for the data in the laminar regime and $n=0.2$ for the data in the turbulent regime. The plots are provided in Fig.\ref{Fig6} and the calibration coefficients in Eq.\eqref{eq8} are provided in Table \ref{Tab1}; these coefficient were obtained with an $R^2$ value of $0.99$ and a confidence level of $95$\%.\

\begin{table}[h!]
\centering
\caption{Calibration coefficients and exponent for laminar and turbulent flow regimes}
\begin{tabular}{|l|c|c|c|}
\hline
\text{Flow regime} &$A$ &$B$ &$n$ \\ \hline
\text{Laminar} &-0.216 &0.798 &$0.4$ \\ \hline
\text{Turbulent} &-50.86 &32.70 &$0.2$ \\ \hline
\end{tabular}
\label{Tab1}
\end{table}

\noindent This calibration channel achieved a maximum pressure difference of 4.95 kPa between adjacent ports with the pump running at maximum capacity, resulting in a maximum shear stress of 12.29 Pa. The range of the calibration channel restricts our results to a single operating point (mass flow rate) because of the large variation in shear stress under the Bernoulli pad.

\section{Bernoulli Pad Experimental Setup}\label{sec4}
\begin{figure}[t!]
\begin{center}
%\psfrag{A}[][]{\small{$H$}}
%\includegraphics[width=0.75\hsize]{Fig7-rev.eps}
\includegraphics[width=0.75\hsize]{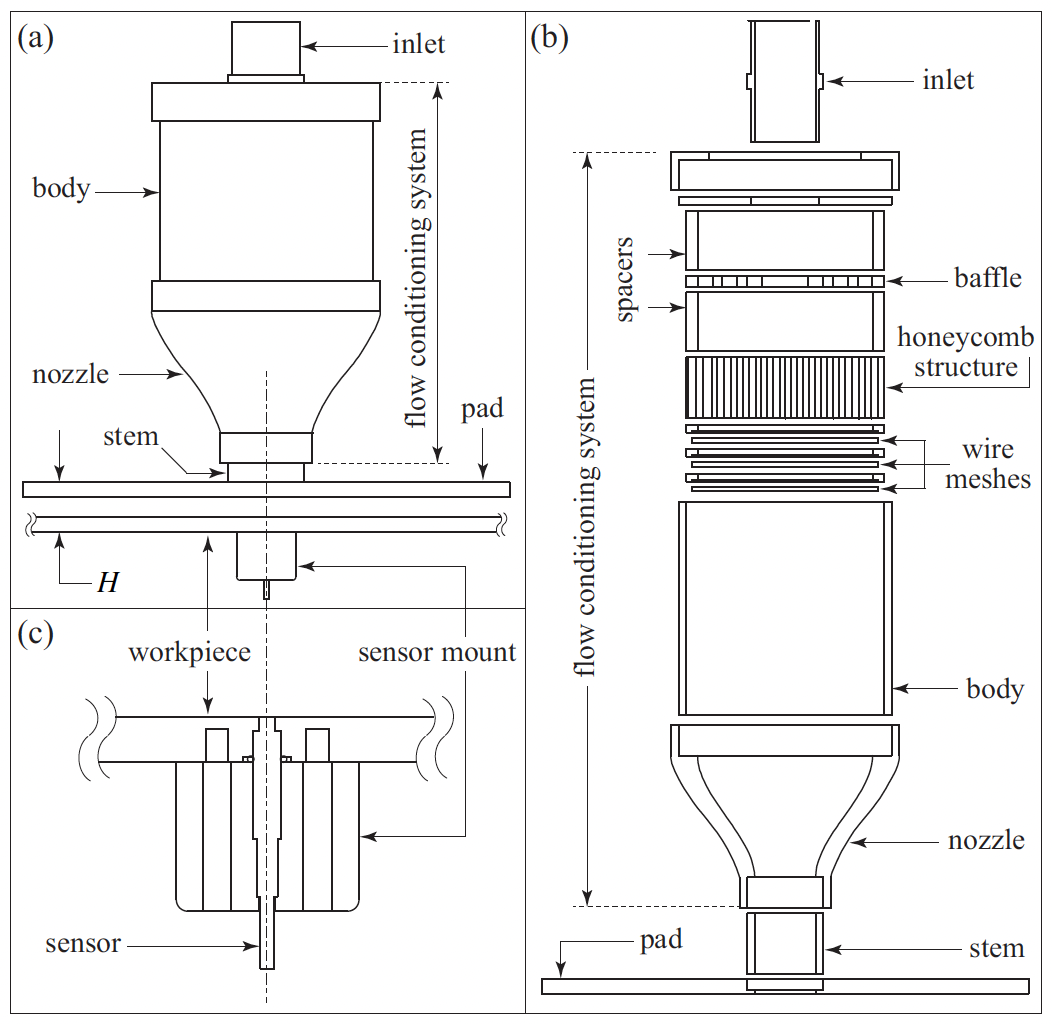}
\end{center}
\caption{\small{(a) Bernoulli pad assembly shown in its nominal configuration over the workpiece (b) exploded view of Bernoulli pad assembly (c) sectional view of flush-mounted sensor in workpiece.}}
\label{Fig7}
\end{figure}

An experimental setup was developed to measure the wall shear stress generated by a Bernoulli pad. An important component of the setup is the Bernoulli pad assembly, comprised of the stem, pad, and flow conditioning section - see Fig.\ref{Fig7}(a) and (b). The stem is a tube with an inside diameter of $d = 25.4$ mm ($1.0$ in), an outside diameter of $31.75$ mm ($1.25$ in), and a length of $88.9$ mm ($3.5$ in). The pad is a circular, flat plate with a diameter of $D = 203.2$ mm ($8.0$ in) and a thickness of $6.35$ mm ($0.25$ in); this thickness is an important dimension only insofar as it prevents the pad from deflecting under fluid loads. A $25.4$ mm ($1.0$ in) hole is located at the center of the pad along with a $31.75$ mm ($1.25$ in) counterbore to $75$\% depth in the pad. The stem interfaces with the pad in the counterbore such that there are negligible steps for the fluid to encounter. Both the stem and the pad are made of cast acrylic, which has sufficient rigidity and low mean surface roughness depth $0.62\,\mu {\rm m} < {\rm Rz} < 0.89\, \mu {\rm m}$ (\cite{bib31}).\

The flow conditioning section is used to obtain a top-hat velocity profile with low-turbulence-intensity at the inlet of the stem. From upstream to downstream, it is composed of the inlet from the pump, a baffle that breaks down the incoming pipe flow, a honeycomb, three wire meshes of decreasing hole size, and a flow contraction - see Fig.\ref{Fig7}(b). A cubic equation was used to design the flow contraction profile (\cite{bib32}). The components of the flow conditioning section were mated with the stem in a similar fashion as the stem-pad interface to avoid disturbances in the flow. Except the inlet, baffle, honeycomb, and wire meshes, the materials of the components of the flow conditioning section were chosen as cast acrylic and polyurethane-coated 3D printed PLA for smooth surface finishes.\
\begin{figure}[t!]
\begin{center}
\includegraphics[width=0.55\hsize]{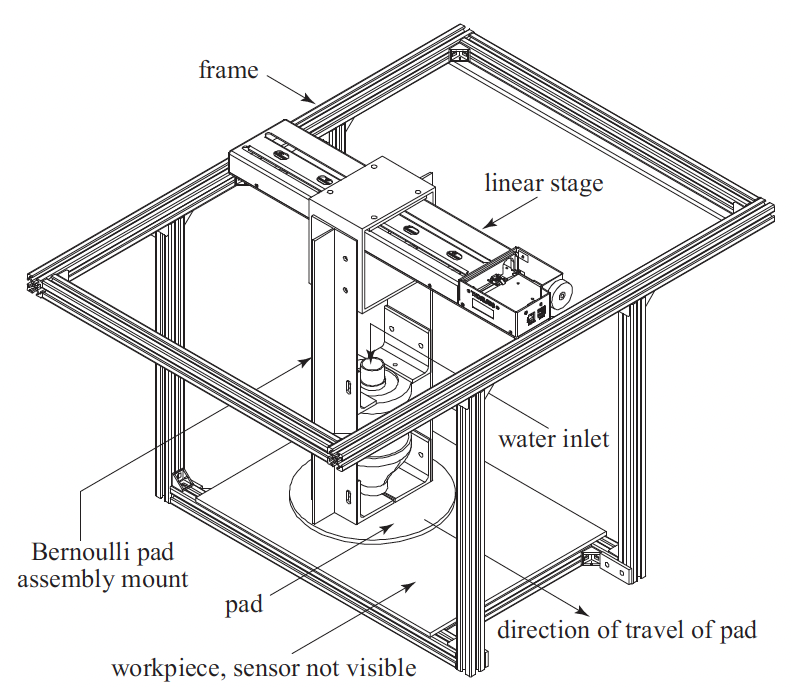}
\end{center}
\caption{\small{An assembled view of the Shear Test Station (STS), comprised of the Bernoulli pad assembly mount, workpiece and linear stage.}}
\label{Fig8}
\end{figure}
 
 The apparatus is also comprised of the workpiece (or wall), sensor mount, linear stage, Bernoulli pad assembly mount and frame. As assembled view of these components, which we will refer to as the Shear Test Station (STS), is shown in Fig.\ref{Fig8}. The workpiece is a rectangular flat plate ($406.4$ mm $\cross$ $304.8$ mm $\cross$ $6.35$ mm) of 7075-T6 aluminum. The surface of the workpiece was sanded and polished to reduce its surface roughness. This process typically results in roughness depth levels $0.5 \,\, \mu{\rm m} < {\rm Rz} < 1.4 \,\, \mu{\rm m}$. The sensor mount is placed in a counterbored hole near the center of the plate which places the sensing surface of the hot-film sensor flush with the workpiece - see Fig.\ref{Fig7}(c). The sensor's grounding lead was permanently fastened to one of the plate's vertical faces. Blind holes allow the sensor mount to be secured to the underside of the workpiece.\

A LTS300 linear translation stage manufactured by THORLABS (\cite {bib33}) was used to move the Bernoulli pad over the length of the workpiece such that the hot-film sensor can measure the wall shear stress along the radial direction. The stage has an absolute accuracy of $50$ $\mu$m  and a repeatability of $2$ $\mu$m. The Bernoulli pad assembly mount, shown in Fig.\ref{Fig8}, is a series of structural aluminum members that interface the Bernoulli pad assembly in Fig.\ref{Fig7}(a) to the linear stage. The joints of the Bernoulli pad assembly mount are comprised of slots and threaded fasteners that allow for setting a uniform gap between the pad and the workpiece and ensure that the pad and the workpiece remain parallel with travel of the linear stage. The fluid power assembly consists of a $250$ W submersible pump, rubber hoses, and a gate valve for flow rate control.\

\section{Wall Shear Stress Measurement}\label{sec5}

\subsection{Procedure}\label{sec51}

\begin{figure*}[t!]
\begin{center}
\includegraphics[width=5.5in]{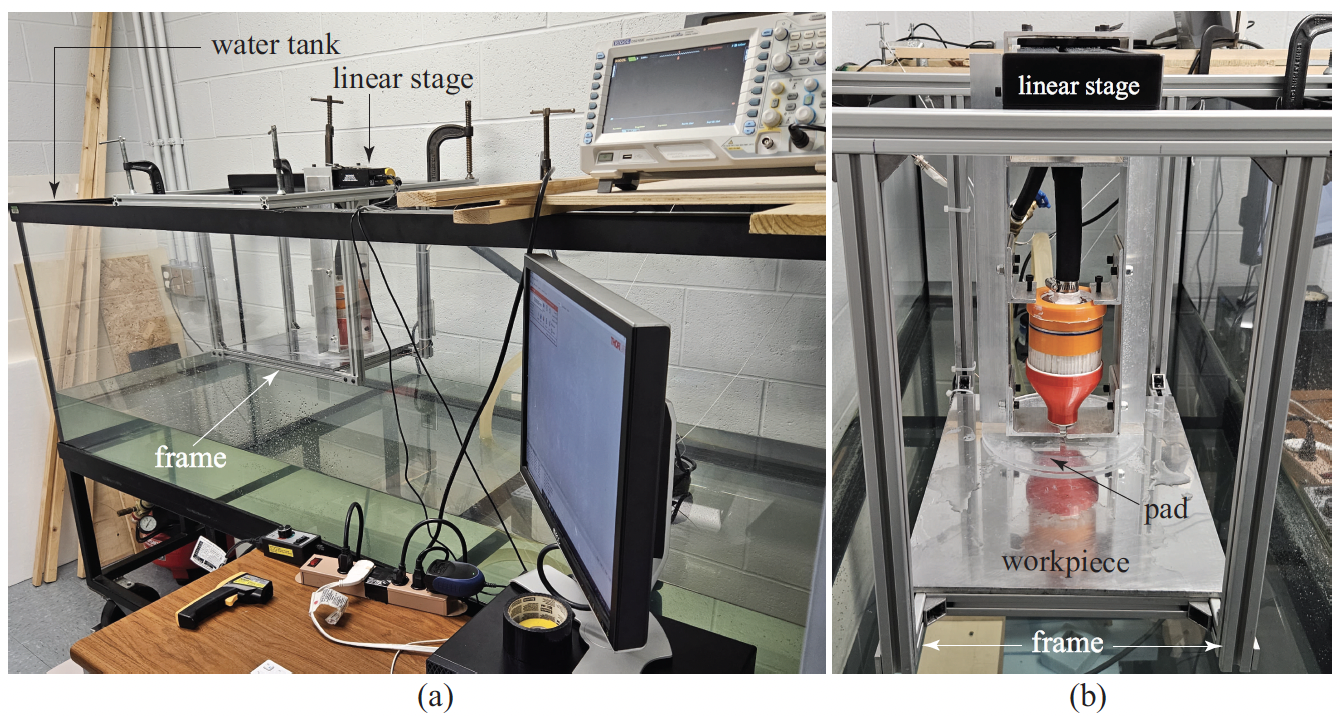}
\end{center}
\caption{\small{(a) A perspective view of the experimental setup (b) a side view showing the STS}}
\label{Fig9}
\end{figure*}

The Shear Test Station (STS) was placed into a water tank with dimensions of $2120$ mm $\cross$ $749$ mm $\cross$ $762$ mm - see Fig.\ref{Fig9}. The frame suspended the workpiece in the center of the tank footprint and $356$ mm from the bottom. The dimensions of the tank and the location of the workpiece were chosen to ensure that the radial outflow between the pad and workpiece was not affected by the water splashing from the tank walls. Shims were placed between the frame and tank to level the STS before it was secured to the tank with clamps. The pump draws water from beneath the workpiece, and this water is at the ambient temperature ($17.25^\circ$ C) before data collection. The density and viscosity of water at this temperature are $\rho = 998.87$ kg/m$^3$ and $\mu = 0.001073$ Pa$\,$s.

In this experiment, the gap height between the pad and the workpiece was chosen to be $H = 1.3$ mm - see Fig.\ref{Fig7} (a). Shims were placed in the gap to ensure that the gap height was uniform; they were removed after ensuring uniform desired gap height. The working mass flow rate was $\dot m = 0.046$ kg/s. This operating point, defined by a combination of $H$ and $\dot m$, was chosen to ensure that the maximum shear stress experienced by the hot-film sensor was within our achievable calibration range\footnote{Higher mass flow rates are also associated with larger maximum local velocities, which produces cavitation near the neck if water is used as the working fluid; this phenomenon is briefly discussed in \cite{bib9}.}. At lower flow rates, the relative error at large $r$ increases significantly, along with a tendency of the flow to become visibly non-axisymmetric.
 
Once the mass flow rate from the pump was at steady state, the linear stage was used to move the Bernoulli pad from its nominal position, and hot-film voltages are measured at each position. At large radii of the Bernoulli pad, we used steps of $5$ mm, which is approximately twice the diameter of the hot-film sensor. The step size was reduced to $1.0$-$2.0$ mm close to the neck of the Bernoulli pad, where large variations in the shear stress are expected. The total distance of travel was approximately $\ell = 83.0$ mm, which ensured that the sensor was never exposed to air.\

The voltage outputs from the hot-film sensor were first corrected for temperature difference between calibration and wall shear stress measurement conditions using (\cite{bib34}):\
\begin{equation*}
E_{\rm corr} = \left(\frac{T_{\rm f} - T_0}{T_{\rm f} - T_{\rm a}}\right)^{0.5} E_{\rm a}
\end{equation*}

\noindent where, $E_{\rm corr}$ is the corrected voltage, $T_0 = 25.49^\circ$C is the water temperature during calibration, $T_{\rm f} = 40^\circ$C is the film temperature, and $T_{\rm a} = 17.25^\circ$C is the water temperature during data acquisition. The corrected voltage values were used to obtain the wall shear stress values using Eq.\eqref{eq8} with the calibration coefficients provided in Table \ref{Tab1}.\

\subsection{Experimental results and comparison with simulation}\label{sec52}

\begin{figure}[t!]
\begin{center}
%\psfrag{A}[][]{\small{$z$}}
%\psfrag{B}[][]{\small{$r$}}
%\psfrag{C}[][]{\small{$D = 203.2$ mm}}
%\psfrag{H}[][]{\small{$d = 25.4$ mm}}
%\psfrag{K}[][]{\footnotesize{workpiece roughness, $\varepsilon_{\rm w}$}}
%\psfrag{P}[][]{\footnotesize{pad roughness, $\varepsilon_{\rm p}$}}
%\psfrag{D}[][]{\small{$H = 1.3$ mm}}
%\psfrag{E}[][]{\small{$\dot m = 0.046$ kg/s}}
%\psfrag{L}[][]{\small{$p_{\rm atm}$}}
%\psfrag{Q}[][]{\small{pad}}
%\psfrag{R}[][]{\small{stem}}
%\psfrag{M}[][]{\small{$\ell= 83.0$ mm}}
%\includegraphics[width=0.75\hsize]{Fig10.eps}
\includegraphics[width=0.75\hsize]{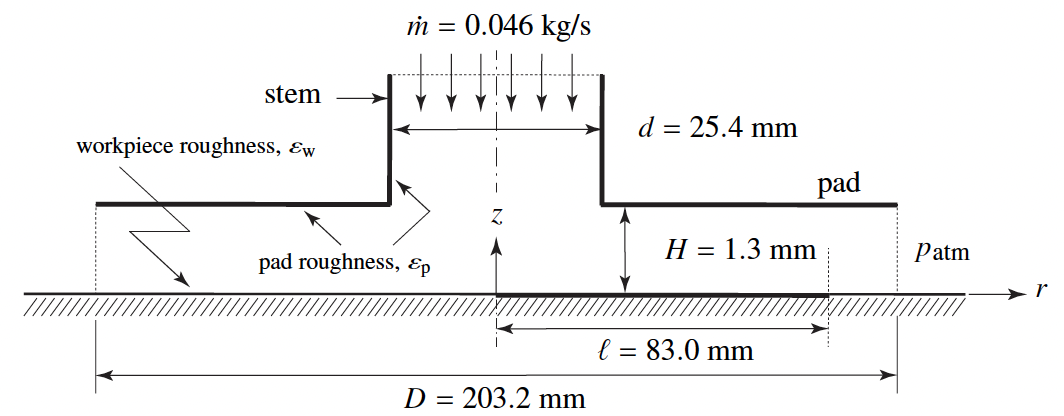}
\end{center}
\caption{A schematic of the Bernoulli pad used in simulations.}
\label{Fig10}
\end{figure}

Numerical simulations were carried out using the Spalart-Allmaras model based on its suitability for adverse pressure gradients (\cite{bib35}) and the Transition-SST model due to its extensive use in our prior work with Bernoulli pads (\cite{bib8, bib9, bib24}). These simulations were carried out for identical flow domain and boundary conditions in the experiments - see Fig.\ref{Fig10}. The domain is axially symmetric without variations in the azimuthal directions. Therefore, we used a two-dimensional axisymmetric model to reduce computational time. Assuming incompressible flow, we imposed a mass flow rate of $\dot{m} = 0.046$ kg/s at the inlet, and exit pressure $p = p_\text{atm}$ at the pad’s outlet. No-slip conditions were imposed on all solid walls. The computational domain is meshed with quadrilateral dominant elements. The conservation equations are solved using SIMPLE algorithm and the pressure and momentum terms are discretized using PRESTO! and second order discretization respectively. The convergence is said to be obtained when the residual is less than $1 \times 10^{-6}$ for all the variables.\

In our experimental setup, we were careful to choose materials and surface finishes which will allow us to describe the pad and workpiece as ``hydraulically smooth". This condition is met when $R_z/\delta_l < 5$ (\cite{bib36}), where $\delta_l$ is the thickness of the laminar sublayer. The representative thickness of our laminar sublayer is $\delta_l = 9.9$ $\mu$m based on the maximum measured value of $\tau_{\rm w}$. Given that $R_z < 1.4$ $\mu$m for polished aluminum and less for cast acrylic, we satisfy the hydraulically smooth descriptor. In simulation, we fix surface roughness for the pad and workpiece of $\varepsilon_\text{p} = \varepsilon_\text{w} = 1$ $\mu$m.\

In our experiments, the hot-film sensor voltage is obtained for the measurement region of length $\ell = 83.0$ mm shown in Fig.\ref{Fig10}. A comparison of these voltages with the voltage values in Fig.\ref{Fig6} indicate that the flow is laminar for $0 \leq r < 12$ and $14 < r \leq 83$ mm and turbulent for $12 \leq r \leq 14$ mm. The change from laminar to turbulent flow close to the neck of the pad is expected. In this region, the mean velocity is high because of the reduced area relative to the stem. This is amplified by a recirculation region resulting from the sharp corner (\cite{bib11}) which further reduces the area through which the flow is moving in the $+r$ direction. The wall shear stress and radial distance are non-dimensionalized using the expressions (\cite{bib4}):
\begin{equation}
\bar \tau_{\rm w} = \left(\frac{\pi \rho d^3}{4 \dot m \mu}\right) \tau_{\rm w}, \qquad \bar r = \left(\frac{2}{D}\right) r
\end{equation}

\begin{figure}[t!]
\begin{center}
%\psfrag{A}[][]{\small{$\bar\tau_{\rm w}$}}
%\psfrag{B}[][]{\small{$\bar r$}}
%\psfrag{C}[][]{\footnotesize{$k$-$\varepsilon$}}
%\psfrag{D}[][]{\footnotesize{$k$-$\omega$}}
%\psfrag{E}[][]{\scriptsize{$\times 10^3$}}
%\includegraphics[width=0.64\hsize]{Fig11-rev2.eps}
\includegraphics[width=0.64\hsize]{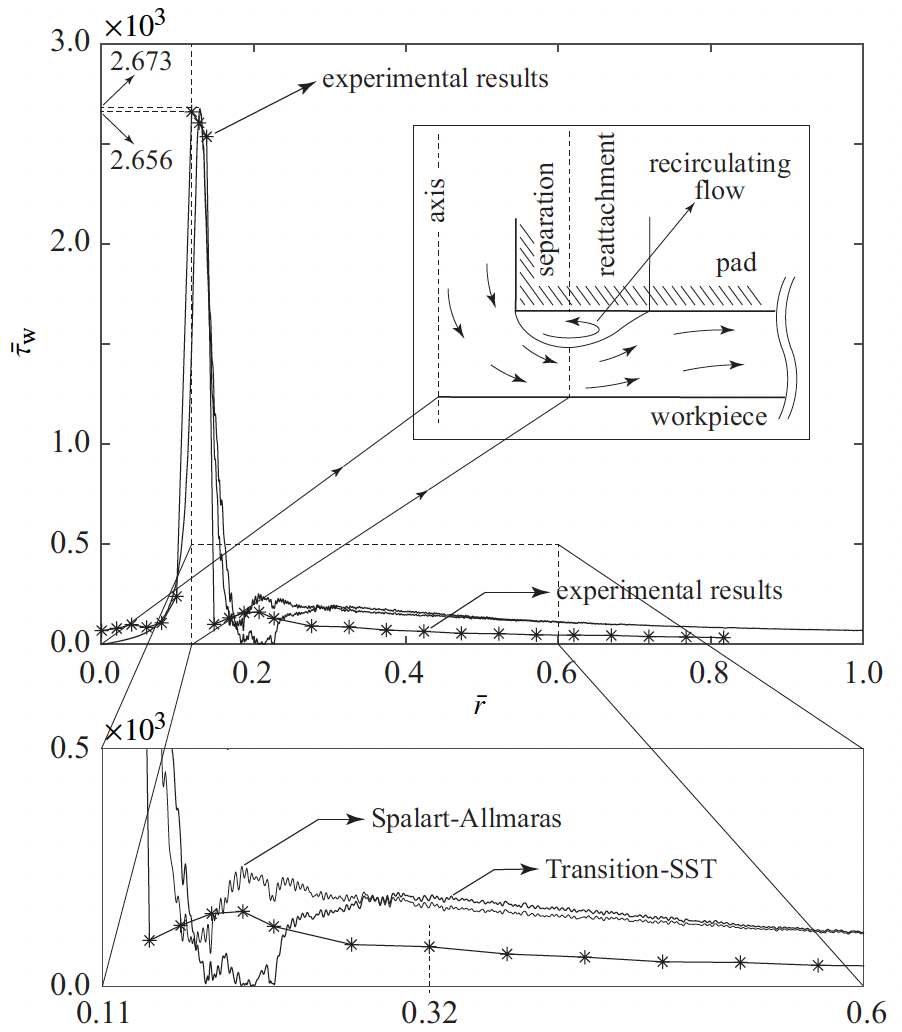}
\end{center}
\caption{\small{Variation of $\bar \tau_{\rm w}$ with $\bar r$: a comparison of simulation and experimental results. The experimental results, which were obtained at discrete values of $\bar r$, are shown using ``$*$" marks. The turbulent calibration coefficients in Table \ref{Tab1} are used from $\bar r \in [0.12, 0.14]$, the laminar values are used in the rest of the domain. The top figure shows the recirculation region and the flow around it (\cite{bib11}) - see figure inset. A magnified view of the dotted portion of the top figure is shown in the bottom figure for comparison of the two turbulence models with experimental results.}}
\label{Fig11} 
\end{figure}

\begin{figure}[t!]
\begin{center}
%\psfrag{A}[][]{\footnotesize{Pa}}
%\psfrag{B}[][]{\footnotesize{$r$ (mm)}}
%\psfrag{C}[][]{\footnotesize{$z$ (mm)}}
%\psfrag{D}[][]{\footnotesize{m/s}}
%\includegraphics[width=0.70\hsize]{Fig12-rev.eps}
\includegraphics[width=0.70\hsize]{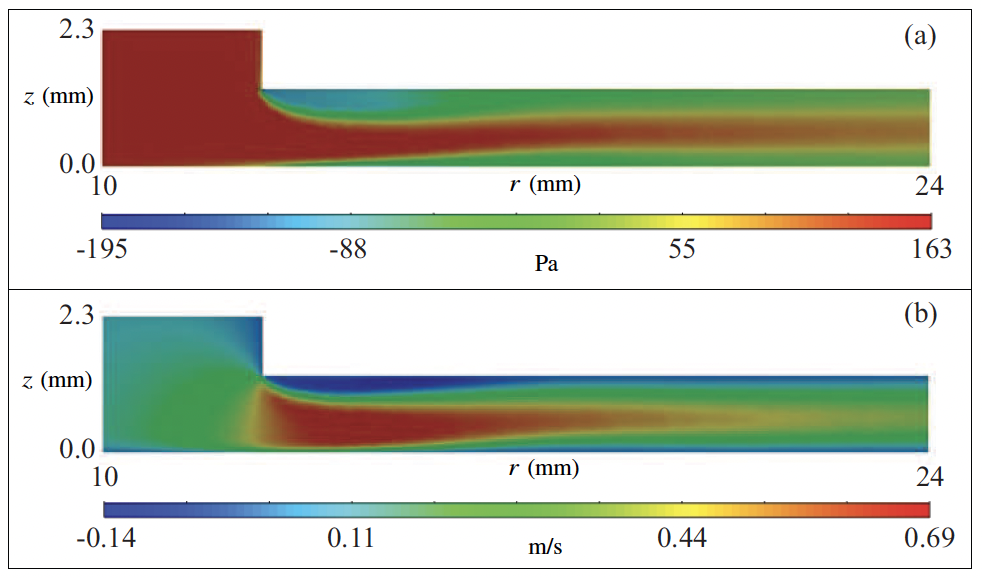}
\end{center}
\caption{\small{Contour plots of (a) total pressure and (b) radial velocity in the computational domain using the Spalart-Allmaras model. The plots are clipped to depict the flow field in the neck region rather than the complete domain.}}
\label{Fig12} 
\end{figure}

The variation in the non-dimensional wall shear stress $\overline{\tau}_\text{w}$ with non-dimensional radial distance $\overline{r}$ is shown in Fig.\ref{Fig11}. Contour plots of the radial velocity and pressure are presented in Fig.\ref{Fig12}. There is a separation bubble attached to the corner of the pad at $r=d/2$. This separation bubble reduces the effective cross-sectional area of the flow field under the pad, increasing the local radial velocity and thereby increasing the shear stress on the workpiece - see the inset of Fig.\ref{Fig11}. As the radial flow moves past the separation bubble, there is momentum transferred from the radial direction to the wall-normal direction, yielding a sharp drop in wall shear stress. The momentum recovers its purely radial direction, and since the boundary layer is still very thin, the shear stress increases even though the mean velocity through the cross section at $r$ is decreasing $\propto r$. However, further downstream, the decreasing mean velocity yields monotonically decreasing wall shear stress. In order to discuss the differences between experiment and simulation, it is helpful to divide the discussion of the data and simulation results into regions of the wall shear stress field.\

\noindent \textbf{Before reattachment ($\bar r \leq 0.14$)}: There is good agreement between both models and experiment on the values of peak shear. This also suggests that the models are accurately capturing the velocities and extent of the recirculation region. From an application perspective, this is promising; the cleaning efficacy of a Bernoulli pad is characterized by the maximum shear it produces (\cite{bib8}), while the deformation and fracture limits on a workpiece are likewise set by the maximum local shear and normal forces. The maximum non-dimensional wall shear stress $\tau_{\rm w, max}$ for all cases, including a power law estimation from (\cite{bib4}), is shown in Table \ref{Tab2}. Inside the core of the impinging jet, the simulations underpredict the experiment. However, this region of the flow is the region which matches the calibration conditions the least well, because of the large gradient in the normal component of velocity. We expect a developing boundary layer in this region and therefore a larger diffusion of thermal energy at a given velocity than in the calibration flow.\
\begin{table*}[t!]
\centering
\caption{Dimensionless maximum wall shear stress values obtained from experiment, numerical models, and power law (\cite{bib4})}
\begin{tabular}{|c||c|c||c|}
\hline
Experiment &Spalart-Almaras &Transition-SST &Power Law \\ \hline
$2656.1$ &$2673.1$ &$2676.3$ &$2548.6$ \\ \hline
\end{tabular}
\label{Tab2}
\end{table*}

\noindent \textbf{Fully radial flow ($ \bar r \geq 0.32$)}: Qualitatively, the gradual decay of wall shear in this region is easily explained by the $1/r$ proportional decay in radial velocity as the cross-sectional area of the gap increases with radius (\cite{bib37}). While the experiments and the simulations both show this behavior, the models substantially over-predict the experiments. In this region of the flow, the experiment predicts that the flow is laminar. In addition to the two models presented in Fig.\ref{Fig11}, we also ran several other turbulence models with no transition model whatsoever, and these all showed over-prediction of the wall shear stress in this region. The tendency to over-predict was also observed in our prior work (\cite{bib4}), where a numerical model was compared to PTV data. We tentatively suggest that the over-prediction in this region is driven by a failure to accurately characterize the flow in this region as laminar, and that better predictions at large $r$ will require a model tuned to capture this phenomenon.\

\noindent \textbf{Near reattachment ($0.14 < \bar r < 0.32$)}: In this region, there is substantial difference between each model and experiment. All realizations indicate a local minimum followed by a net rise before $r=0.3$. The experiment shows an immediate rise to a local maximum near $r=0.2$. The Spalart-Allmaras model was chosen as a candidate model because it was designed to perform well in wall-bounded flows at moderate to low Reynolds numbers under adverse pressure gradients (\cite{bib35}). The Spalart-Allmaras model indicates a similar local minimum, slightly downstream of experiment. Likewise, the local maximum is further downstream than observed experimentally, consistent with a separation bubble which is larger in the streamwise direction. The four-equation Transition SST model shows a broad region of near-zero shear stress followed by a rise to $r=0.3$. The $\tau_w\approx0$ region is likely an indicator of laminar separation (\cite{bib4}), followed by another transition to turbulence. The cited work found Transition SST to perform well in a fully separated flow. However, there are empirical correlations used in the development of the Transition SST model (\cite{bib39}) which were performed on a flat plate. It may be that the current flow field, characterized by sharp radial pressure gradients, presents an out-of-sample challenge for the correlations.\

\section{Conclusion}\label{sec6}

The experimental work presented here uses a constant temperature anemometer with a hot-film sensor to quantify the wall shear stress generated by the action of Bernoulli pad over a proximally located workpiece. An experimental setup, consisting of a rectangular channel, is designed to calibrate the wall shear stress. The calibration of the sensor is carried out separately for laminar and turbulent regimes. These calibration relations are subsequently used to measure the wall shear stress generated by a Bernoulli pad. It should be mentioned that this experimental effort, which quantifies the wall shear stress generated by a Bernoulli pad with water as the working fluid, is the first of its kind.\

The numerical simulations accurately predict the maximum shear stress, and qualitatively correct behavior, including a secondary peak in the shear stress associated with streamline reattachment and shear stress declining approximately $\propto 1/r$ thereafter. The position of the maximum wall shear stress is found to be very close to the neck of the Bernoulli pad, right below the belly of the recirculation region. The RANS computational models we used do not accurately predict the magnitude of the secondary peak, and significantly overpredict the shear stress at large $r$.\

A better turbulence scheme, such as LES or DNS, may provide a better match for the entire domain of $\overline{r}$; however, this will require significantly higher computational effort and lies in the scope of future research. Because the most important aspect of the shear produced by a  Bernoulli pad for grooming and cleaning applications is its maximum value, we suggest that simpler and faster computational tools are adequate for this purpose. Modeling improvements to better capture the shear stress behavior at all $r$ should be focused on relaminarization and reattachment.\

\begin{Backmatter}

% \paragraph{Acknowledgements}
% We gratefully acknowledge the advice of John Smith who
% commented on a version of this manuscript.

\paragraph{Funding Statement}
This work was partially funded by the Office of Naval Research, grant no. N00014220-1-2170, through a subcontract provided by the University of Massachusetts, Dartmouth.

\paragraph{Declaration of Interests}
The authors declare no conflict of interest.

% \paragraph{Author Contributions}
% R.B. and J.G.S. created the research plan, designed
% experiments, and formulated analytical problem. R.B. led
% model solution and performed all experiments. R.B. and
% J.G.S. wrote the manuscript.

\paragraph{Data Availability Statement}
Raw data are available from the corresponding author.

\paragraph{Ethical Standards}
The research meets all ethical guidelines, including adherence to the legal requirements of the study country.

% \paragraph{Supplementary Material}
% Methods section and Supplementary information are available
% at \url{https://doi.org/10.1017/flo.2021.1}.

\end{Backmatter}

\end{document}